\documentclass[12pt,epsf,amssymb]{article} \textheight =23 truecm
\def\lsim{\raise0.3ex\hbox{$<$\kern-0.75em\raise-1.1ex\hbox{$\sim$}}}
\def\gsim{\raise0.3ex\hbox{$>$\kern-0.75em\raise-1.1ex\hbox{$\sim$}}}
  \def\bea{\begin{eqnarray}}
\def\eea{\end{eqnarray}} \def\beq{\begin{equation}}
\def\eeq{\end{equation}} 
\def\beeq{\begin{eqnarray}} \def\eeeq{\end{eqnarray}} \def\R{ {\rm R
\kern -.31cm I \kern .15cm}} \def\C{ {\rm C \kern -.15cm \vrule
width.5pt \kern .12cm}} \def\Z{ {\rm Z \kern -.27cm \angle \kern
.02cm}} \def\N{ {\rm N \kern -.26cm \vrule width.4pt \kern .10cm}}
\def\1{{\rm 1\mskip-4.5mu l} }
\usepackage{amsmath}\usepackage{amssymb}

\usepackage{graphics} \usepackage{graphicx} \usepackage{epsfig}

\begin{document} \begin{center} 

{\large \bf Isgur-Wise functions, Bjorken-Uraltsev Sum Rules \\ and their Lorentz group interpretation \\\ \vskip 3 truemm {\it In Memoriam} Nikolai Uraltsev} 
\par \vskip 1. truecm

 {\bf L. Oliver \footnote{Luis.Oliver@th.u-psud.fr} and J.-C. Raynal}
\par \vskip 4 truemm

{\it Laboratoire de Physique Th\'eorique}\footnote{Unit\'e Mixte de
Recherche UMR 8627 - CNRS }\\    {\it Universit\'e de Paris XI,
B\^atiment 210, 91405 Orsay Cedex, France} 

\end{center}
\par \vskip 8 truemm

\begin{abstract}
In the heavy quark limit of QCD, using the Operator Product Expansion and the non-forward amplitude, as proposed by Nikolai Uraltsev, we formulate sum rules that generalize Bjorken and Uraltsev sum rules. We recover the Uraltsev lower bound for the slope of the Isgur-Wise (IW) function, that we generalize to higher derivatives. We show that these results have a clear interpretation in terms of the Lorentz group, since the IW function is given by an overlap between the initial and final light clouds, related by Lorentz transformations. Both the Lorentz group and the Sum Rules approach are equivalent. Moreover, we formulate an integral representation of the IW function with a positive measure. Inverting this integral formula, we obtain the measure in terms of the IW function, allowing to formulate criteria to decide if a given ansatz for the IW function is compatible or not with the sum rule constraints. We compare these theoretical constraints to some forms proposed in the literature.

\end{abstract}

 \vskip 5 truemm

LPT-Orsay-14-06

\section{Introduction}

Nikolai Uraltsev made very important contributions to Heavy Quark Effective Theory, concerning both the inclusive $\overline{B} \to X_c \ell \nu$ and exclusive processes $\overline{B} \to D(D^*) \ell \nu$. We will be here concerned with his work on exclusive $B$ decays.\par 
In the heavy quark limit \cite{IW-1}, Bjorken formulated a Sum Rule (SR) \cite{BJORKEN} relating the slope $\rho^2 = -\xi'(1)$ of the elastic IW function $\xi(w)$ to the IW functions at zero recoil for the inelastic transitions changing parity, namely $\tau_{1/2}^{(n)}(w)$, $\tau_{3/2}^{(n)}(w)$, corresponding to transitions of the type ${1 \over 2}^- \to {1 \over 2}^+, {3 \over 2}^+$ \cite{IW-2}. In the notation of \cite{IW-2}, the SR writes
\begin{equation}
\rho^2 = {1 \over 4} + \sum_n \left[|\tau_{1/2}^{(n)}(1)|^2+2|\tau_{3/2}^{(n)}(1)|^2\right]\,.
\label{ra_eq1}
\end{equation}

\noindent This SR implies the famous Bjorken's lower bound on the slope 
\begin{equation}
\rho^2 \geq {1 \over 4}\,.
\label{ra_eq2}
\end{equation}

A decade later, came as a great surprise a new SR formulated by N. Uraltsev \cite{URALTSEV-1} involving the same inelastic IW functions :
\begin{equation}
\sum_n \left[|\tau_{3/2}^{(n)}(1)|^2-|\tau_{1/2}^{(n)}(1)|^2\right] = {1 \over 4}\,.
\label{ra_eq3}
\end{equation}

\noindent The combination of both SR (\ref{ra_eq1})(\ref{ra_eq3}) implies the much stronger lower bound
\begin{equation}
\rho^2 \geq {3 \over 4}\,.
\label{ra_eq4}
\end{equation}

The technique used by Uraltsev to obtain the SR (\ref{ra_eq3}) was quite original. He considered the $T$-product of two currents away from the forward direction :
\begin{equation}
i\int d^4x e^{-iq.x} < \overline{B}(v_f)|T[J(x)J^+(0)]|\overline{B}(v_i) >, \qquad \qquad J = \overline{c} \Gamma b\,,
\label{ra_eq5}
\end{equation}

\noindent where $v_i \not = v_f$, and performed the $1/m_Q$ expansion to obtain the corresponding Operator Product Expansion (OPE) in the direct channel. In other words, he considered non-forward transitions of the type $\overline{B}(v_i) \to D^{(n)}(v') \to \overline{B}(v_f)$, with $v_i \not = v_f$ and $v'$ being the intermediate hadron four-velocity.

By the way, the bound (\ref{ra_eq4}) was obtained earlier in a class of quark models \cite{BT} that yield covariant current matrix elements in the heavy quark limit \cite{LOPR}. Later on we did realize that this was the case because, in the heavy quark limit, this class of models satisfy Isgur-Wise scaling and also both Bjorken and Uraltsev SR \cite{MLOPR-1,LOPRM}.\par
Other important results in this field were obtained by N. Uraltsev, in particular the study of the limit in which the elastic slope reaches its lower limit : 
\begin{equation}
\rho^2 \to {3 \over 4}\,,
\label{ra_eq6}
\end{equation}

\noindent called by Uraltsev $BPS\ limit$ \cite{URALTSEV-2}, on which we will come back below.

Our contribution to this topic began by trying to understand Uraltsev's results and generalize them to higher derivatives of the IW function $\xi(w)$.

\section{Non-forward direction Sum Rules in the heavy quark limit}

Proceeding like N. Uraltsev, we did consider the non-forward amplitude (\ref{ra_eq5}) in a covariant way, and obtained SR that have the general concise form \cite{LOR-1}
\begin{equation}
L_{Hadrons} (w_i, w_f, w_{if}) = R_{OPE} (w_i, w_f, w_{if})\,,
\label{ra_eq7}
\end{equation}

\noindent where 
\begin{equation}
w_i = v_i.v'\,, \qquad \qquad w_f = v_f.v'\,, \qquad \qquad w_{if} = v_i.v_f\,,
\label{ra_eq8}
\end{equation}
\noindent i.e. the consideration of the non-forward amplitude means to extend the $OPE$ formula to $w_{if} \not = 1$, while Bjorken's SR was obtained for $w_{if} = 1$.\par 
The l.h.s. of the SR (\ref{ra_eq7}) $L_{Hadrons} (w_i, w_f, w_{if})$ represents the sum over the intermediate $D^{(n)}(v')$ states while the r.h.s $R_{OPE} (w_i, w_f, w_{if})$ corresponds to the OPE counterpart.\par 

The domain of the variables (\ref{ra_eq8}) is as follows
$$w_i \geq 1, \qquad\qquad w_f \geq 1\,,$$
\begin{equation}
w_iw_f - \sqrt{(w_i^2-1)(w_f^2-1)} \leq w_{if} \leq w_iw_f + \sqrt{(w_i^2 - 1) (w_f^2 -1)}\,,
\label{ra_eq9}
\end{equation}

\noindent and there is the sub-domain:
$$w_i = w_f = w\,,$$
\begin{equation}
w \geq 1\,, \qquad\qquad 1 \leq w_{if} \leq 2w^2-1\,.
\label{ra_eq10}
\end{equation}

We make somewhat more explicit the SR (\ref{ra_eq7}) considering currents of the form ($\Gamma_i$, $\Gamma_f$ are Dirac matrices) 
\begin{equation}
\label{ra_eq11}
\overline{h}_{v'} \Gamma_i h_{v_i}\,, \qquad \qquad \overline{h}_{v_f} \Gamma_f h_{v'}\,,  
\end{equation}

\noindent and one gets :
$$\sum_{D^{(n)}} <\overline{B}_f (v_f)|\Gamma_f|D^{(n)}(v')> <D^{(n)} (v')|\Gamma_i|\overline{B}_i(v_i)> \xi^{(n)}(w_i) \xi^{(n)}(w_f)$$
\begin{equation}
+\ {\rm Other\ excited\ states} = - 2 \xi (w_{if}) \ <\overline{B}_f (v_f)|\Gamma_f P'_+ \Gamma_i|B_i(v_i)>\,,
\label{ra_eq12}
\end{equation}

\noindent where the ground state ${1 \over 2}^-$ and its radial excitations, together with their IW functions $\xi(w)$, $\xi^{(n)}(w)\ (n \not = 0)$ are made explicit and $P'_+ = \displaystyle{{1 +{/\hskip -1.5 truemm v}' \over 2}}$ is the positive energy projector over the intermediate heavy quark $c$.

\section{Generalized Isgur-Wise functions and generalized Sum Rules}

We now consider higher excited states with a light cloud angular momentum $j$ and spin $J$ and transitions between the $\overline{B}$,  pseudoscalar ground state with $(j^P, J^P) = \left ( {1 \over 2}^-, 0^-\right )$ to the whole tower of excited states \cite{FALK}
$D^{(n)}$ with $(j^P, J^P), J = j \pm {1 \over 2}, j = L \pm {1 \over 2}, P = (-1)^{L+1}$. One gets two independent Sum Rules.\par 
Choosing for example  
\begin{equation}
\Gamma_i = {/\hskip-2 truemm v}_i\,, \qquad \qquad \qquad \Gamma_f = {/\hskip-2 truemm v}_f\,,
\label{ra_eq13}
\end{equation}

\noindent gives the $Vector\ Sum\ Rule$ :
$$(w+1)^2 \sum_{L\geq 0} {L+1 \over 2L+1}\ S_L (w,w_{if}) \sum_n \left [ \tau_{L+1/2}^{(L)(n)}(w)\right ]^2 +\ \sum_{L\geq 1}  S_L (w,w_{if}) \sum_n \left [ \tau_{L-1/2}^{(L)(n)}(w)\right ]^2$$
\begin{equation}
= \left ( 1 + 2w + w_{if}\right ) \xi (w_{if})\,,
\label{ra_eq14}
\end{equation}

\noindent while choosing  
\begin{equation}
\Gamma_i = {/\hskip-2 truemm v}_i\gamma_5\,, \qquad \qquad \qquad \Gamma_f = {/\hskip-2 truemm v}_f\gamma_5\,,
\label{ra_eq15}
\end{equation}

\noindent one obtains the $Axial\ Sum\ Rule$ :
$$\sum_{L\geq 0}  S_{L+1} (w,w_{if}) \sum_n \left [ \tau_{L+1/2}^{(L)(n)}(w)\right ]^2 
 +\ (w-1)^2 \sum_{L\geq 1}  {L \over 2L-1} S_{L-1} (w,w_{if}) \sum_n \left [ \tau_{L-1/2}^{(L)(n)}(w)\right ]^2$$
\begin{equation}
= -  \left ( 1 - 2w + w_{if}\right ) \xi (w_{if})\,,
\label{ra_eq16}
\end{equation}
 
\noindent where $\tau_{L\pm 1/2}^{(L)(n)}(w)$ are the IW functions corresponding to the transitions ${1 \over 2}^- \to\left (L \pm {1 \over 2} \right )^P,\  P = (-1)^{L+1}$, the function $S_{L} (w,w_{if})$ is a Legendre polynomial \cite{LOR-1} :
\begin{equation}
S_L(w, w_{if}) = \sum_{0 \leq k \leq L/2} C_{L,k}\left ( w^2-1\right )^{2k} \left (w^2 - w_{if}\right )^{L-2k}\,,
\label{ra_eq17}
\end{equation}

\noindent and the coefficient $C_{L,k}$ is given by : 
\begin{equation}
C_{L,k} = (-1)^k {(L!)^2 \over (2L)!}\ {(2L-2k)! \over k!(L-k)!(L-2k)!}\,.
\label{ra_eq18}
\end{equation}

The function (\ref{ra_eq17}) is defined as
\begin{equation}
S_L(w_i, w_f, w_{if}) = v_{f\nu_1}...v_{f\nu_L} T^{v_{f\nu_1}...v_{f\nu_L}
,v_{i\mu_1}...v_{i\mu_L}}v_{i\mu_1}...v_{i\mu_L}\,,
\label{ra_eq19}
\end{equation} 

\noindent for $w_i = w_f = w$. The object $ T^{v_{f\nu_1}...v_{f\nu_L},v_{i\mu_1}...v_{i\mu_L}}$ is the projector on the polarization tensor of integer spin L :
\begin{equation}
T^{v_{f\nu_1}...v_{f\nu_L},v_{i\mu_1}...v_{i\mu_L}} = \sum_{\lambda} \epsilon'^{(\lambda)*{\nu_1}...{\nu_L}}\epsilon'^{(\lambda){\mu_1}...{\mu_L}}\,,
\label{ra_eq20}
\end{equation}

\noindent and depends on the intermediate velocity $v'$. The polarization tensor $\epsilon'^{(\lambda){\mu_1}...{\mu_L}}$ is symmetric, traceless and transverse, $g_{\mu_i\mu_j}\epsilon'^{(\lambda){\mu_1}...{\mu_L}} = v'_{\mu_i} \epsilon'^{(\lambda){\mu_1}...{\mu_L}} = 0$. Although (\ref{ra_eq20}) is explicitly very complicated, its contraction (\ref{ra_eq19}) can be computed following the method exposed in ref \cite{LOR-1}. \par
Equations (\ref{ra_eq14}) and (\ref{ra_eq16}) are two independent SR of the general form (\ref{ra_eq7}). Differentiating these Sum Rules within the domain (\ref{ra_eq10}) and going to the corner of the domain $w_{if} \to 1, w \to 1$
\begin{equation}
\left [{d^{p+q} (L_{Hadrons}-R_{OPE}) \over dw_{if}^p dw^q} \right ]_{w_{if}=w=1} = 0\,,
\label{ra_eq21}
\end{equation}

\noindent one gets a whole tower of sum rules.\par

At zero recoil, summing the Vector and the Axial SR one finds
\begin{equation}
\xi^{(L)}(1) 
= {1 \over 4} (-1)^L L! \sum_n \left[{L+1 \over 2L+1} 4[\tau^{(L)(n)}_{L+1/2}(1)]^2 + [\tau^{(L-1)(n)}_{L-1/2}(1)]^2 + [\tau^{(L)(n)}_{L-1/2}(1)\right]^2\,,
\label{ra_eq22}
\end{equation}

\noindent that is a generalization of Bjorken SR (\ref{ra_eq1}), that corresponds to $L = 1$.\par
On the other hand, combining also the Axial SR and (\ref{ra_eq22}) one gets
\begin{equation}
\sum_n \left[{L \over 2L+1} [\tau^{(L)(n)}_{L+1/2}(1)]^2 - {1 \over 4}  [\tau^{(L)(n)}_{L-1/2}(1)]^2\right] = \sum_n {1 \over 4} [\tau^{(L-1)(n)}_{L-1/2}(1)]^2\,,
\label{ra_eq23}
\end{equation}

\noindent the generalization of Uraltsev SR (\ref{ra_eq3}), that is obtained for $L = 1$. 

From these SR one finds strong constraints on the derivatives of $\xi(w)$. In particular, one finds
\begin{equation}
(-1)^L \xi^{(L)}(1) \geq {(2L+1)!! \over 2^{2L}}\,,
\label{ra_eq24}
\end{equation}

\noindent that reduces to the bound (\ref{ra_eq4}) for $L = 1$, and generalizes it for all $L$.\par
From careful examination of the several equations obtained from (\ref{ra_eq14}) and (\ref{ra_eq16}) one obtains the improved bound on the curvature in terms of the slope \cite{LOR-2} : 
\begin{equation}
\xi''(1) \geq {1 \over 5} \left [ 4 \rho^2 + 3(\rho^2)^2 \right ]\,.
\label{ra_eq25}
\end{equation}

\noindent The QCD radiative corrections to this relation have been carefully studied by M. Dorsten \cite{DORSTEN}.

\section{The BPS limit of HQET} 

Uraltsev \cite{URALTSEV-2} made a very interesting observation by the consideration of the matrix elements of dimension 5 operators in HQET, the kinetic operator and the chromomagnetic operator :
\begin{equation}
\mu_{\pi}^2 = - {1 \over 2m_B} < \overline{B}|\overline{h}_v (iD)^2 h_v|\overline{B} >\,,
\label{ra_eq26}
\end{equation}
\begin{equation}
\mu_G^2 = {1 \over 2m_B} < \overline{B}|{g_s \over 2} \overline{h}_v \sigma_{\alpha \beta}G^{\alpha \beta} h_v|\overline{B} >\,.
\label{ra_eq27}
\end{equation}

These matrix elements are given in terms of ${1 \over 2}^- \to {1 \over 2}^+, {3 \over 2}^+$ IW functions $\tau_j^{(n)}$ and level spacings $\Delta E_j^{(n)}$, as obtained from the OPE by I. Bigi, M. Shifman, N. Uraltsev and A. Vainshtein \cite{BIGI} :
\begin{equation}
\mu_{\pi}^2 = 6 \sum_n\ [\Delta E_{3/2}^{(n)}]^2 [\tau^{(n)}_{3/2}(1)]^2 + 3 \sum_n\ [\Delta E_{1/2}^{(n)}]^2 [\tau^{(n)}_{1/2}(1)]^2\,,
\label{ra_eq28}
\end{equation}
\begin{equation}
\mu_G^2 = 6 \sum_n\ [\Delta E_{3/2}^{(n)}]^2 [\tau^{(n)}_{3/2}(1)]^2 - 6 \sum_n\ [\Delta E_{1/2}^{(n)}]^2 [\tau^{(n)}_{1/2}(1)]^2\,.
\label{ra_eq29}
\end{equation}

\noindent These expressions imply the inequality $\mu_{\pi}^2 \geq \mu_G^2$. From the fit to the inclusive semileptonic decay rate one gets $\mu_{\pi}^2 \cong 0.40\ {\rm GeV}^2$, while from the meson hyperfine splitting one obtains $\mu_G^2 \cong 0.35\ {\rm GeV}^2$.\par
From these numerical values, N. Uraltsev \cite{URALTSEV-2} assumed the following limit
\begin{equation}
\mu_{\pi}^2 = \mu_G^2 \qquad \to \qquad \tau^{(n)}_{1/2}(1) = 0\,,
\label{ra_eq30}
\end{equation}

\noindent that he called {\it BPS limit} of HQET. N. Uraltsev made the observation that, in this limit, the slope of the elastic IW function $\xi(w)$ reaches its lower bound (\ref{ra_eq6}) $\rho^2 \to {3 \over 4}$, as can be seen easily from the SR written above and using (\ref{ra_eq30}).\par
Generalizing Uraltsev's arguments we did found the condition on the curvature \cite{JLOR}
\begin{equation}
\tau^{(2)(n)}_{3/2}(1) = 0 \qquad \to \qquad \xi"(1) = {15 \over 16}\,. 
\label{ra_eq31}
\end{equation}

Moreover, using the whole tower of SR formulated above we did demonstrated by induction that in the BPS limit one obtains \cite{JLOR}
\begin{equation}
\tau^{(L)(n)}_{L-1/2}(1) = 0 \qquad \to \qquad (-1)^L\xi^{(L)}(1) = {(2L+1)!! \over 2^{2L}}\,.
\label{ra_eq32}
\end{equation}

Assuming reasonable continuity regularities this implies the following explicit expression for the elastic IW function in the BPS limit : 
\begin{equation}
\xi(w) = \left({2 \over w+1}\right)^{3/2}\,.
\label{ra_eq33}
\end{equation}

As we have demonstrated elsewhere \cite{LOR-4} and will expose below, this simple fully explicit form has a transparent group theoretical interpretation in terms of the Lorentz group. 

\section{The Lorentz group  and the heavy quark limit of QCD}

Hadrons with one heavy quark such that $m_Q >> \Lambda_{QCD}$ can be thought as a bound state of a light cloud in the color source of the heavy quark. Due to its heavy mass, the latter is unaffected by the interaction with soft gluons.\par
In this approximation, the decay of a heavy hadron with four-velocity $v$ into another hadron with velocity $v'$, for example the semileptonic decays $\overline{B} \to D^{(*)}\ell \overline{\nu}_{\ell}$ or $\Lambda_b \to \Lambda_c \ell \overline{\nu}_{\ell}$, occurs just by free heavy quark decay produced by a current, and rearrangement of the light cloud to follow the heavy quark in the final state and constitute the final heavy hadron.\par
The dynamics is contained in the complicated light cloud, that concerns long distance QCD and is not calculable from first principles. Therefore, one needs to parametrize this physics through form factors, the IW functions.\par
The matrix element of a current between heavy hadrons containing heavy quarks $Q$ and $Q'$ can thus be factorized as follows \cite{FALK,LOR-3}
$$<H'(v'), J'\ m'|J^{Q'Q}(q)|H(v), J\ m >\ =\ < {1 \over 2}\ \mu', j' M'| J' m' >< {1 \over 2}\ \mu, j M| J m >$$
\begin{equation}
\times <Q'(v'), {1 \over 2}\ \mu'|J^{Q'Q}(q)|Q(v),{1 \over 2}\ \mu > < {\rm cloud},v',j',M'|{\rm cloud},v,j,M>\,,
\label{ra_eq34}
\end{equation} 

\noindent where $v$, $v'$ are the initial and final four-velocities, and $j$, $j'$, $M$, $M'$ are the angular momenta and corresponding projections of the initial and final light clouds. \par
The current affects only the heavy quark, and all the soft dynamics is contained in the {\it overlap} between the initial and final light clouds $<v',j',M'|v,j,M>$, that follow the heavy quarks with the same four-velocity. This overlap is independent of the current heavy quark matrix element, and depends on the four-velocities $v$ and $v'$. The IW functions are given by these light clouds overlaps.\par
An important hypothesis has been done in writing the previous expression, namely neglecting {\it hard gluon radiative corrections}.\par
As we will make explicit below, the light cloud belongs to a Hilbert space, and transforms according to a unitary representation of the Lorentz group. Then, as we have shown\cite{LOR-3}, the whole problem of getting rigorous constraints on the IW functions amounts to decompose unitary representations of the Lorentz group into irreducible ones. This allows to obtain, for the IW functions, general integral formulas in which the crucial point is that {\it  the measures are positive}.\par
In \cite{LOR-3} we did treat the case of a light cloud with angular momentum $j = 0$ in the initial and final states, as happens in the baryon semileptonic decay $\Lambda_b \to \Lambda_c \ell \overline{\nu}_{\ell}$. \par
The sum rule method exposed above is completely equivalent to the method based on the Lorentz group, as demonstrated in ref \cite{LOR-3}.\par
Ignoring spin complications, the IW function writes then simply (e.g. in the special case $j = 0$) : 
\begin{equation}
\xi(v.v') =\ <U(B_{v'})\phi_0|U(B_v)\phi_0>\,,
\label{ra_eq35}
\end{equation} 

\noindent where $B_v$ and $B_{v'}$ are the corresponding boosts.\par

One can easily get a physical picture of why the Lorentz group plays an essential role as far as the Isgur-Wise function is concerned. In the limit in which factorization holds, i.e. switching off the hard gluon radiative corrections between the heavy quark and the light cloud, the heavy quark of initial velocity $v$ is strucked by a weak current - a hard process - and the deviated final heavy quark gets a different velocity $v'$. The factorization means that, {\it because of confinement} - a soft process -, the light cloud {\it unchanged} state, after the heavy quark is strucked by the current, must be expanded in terms of eigenstates of the Hamiltonian corresponding to the four-velocity $v'$. This picture corresponds to the so-called "approximation soudaine" \cite{MESSIAH}. Therefore, the overlap of the light cloud before and after the interaction with the current, $< {\rm cloud},v'|{\rm cloud},v>$, is constrained only by its quantum numbers and by kinematics.\par
The crucial point is that {\it the states of the light component make up a Hilbert space in which acts a unitary representation of the Lorentz group}. In fact, this is more or less implicitly stated, and used in the literature\cite{FALK}.\par

To see the point more clearly, let us go into the physical picture which is at the basis of (\ref{ra_eq35}). Considering first a heavy hadron {\it at rest}, with velocity $v_0 = (1,0,0,0)$ its light component is submitted to the interactions between the light particles, light quarks, light antiquarks and gluons, and to the external chromo-electric field generated by the heavy quarks at rest. This chromo-electric field does not depend on the spin $\mu$ of the heavy quark nor on its mass. We shall then have a complete orthonormal system of energy eigenstates $|v_0,j,M>$ of the light component, where $j$ and $M$ are the angular momentum quantum numbers, $<v_0,j',M',\alpha'|v_0,j,M>\ = \delta_{j,j'} \delta_{M,M'}$ 

Now, for a heavy hadron moving with a velocity $v$, the only thing which changes for the light component is that the external chromo-electric field generated by the heavy quark at rest is replaced by a chromo-electromagnetic field generated by the heavy quark moving with velocity $v$. Neither the Hilbert space describing the possible states of the light component, nor the interactions between the light particles, are changed. We shall then have {\it a new complete orthonormal system} of energy eigenstates $|v,j,M>$ in the same Hilbert space. Then, because the colour fields generated by a heavy quark for different velocities are related by Lorentz transformations, we may expect that the energy eigenstates of the light component will, for various velocities, be themselves related by Lorentz transformations acting in their Hilbert space.\par

\section{From a Lorentz representation to Isgur-Wise functions}

For half-integer spin $j$, as is the case of the ground state mesons $j^P = {1 \over 2}^-$, the polarization tensor is a Rarita-Schwinger tensor-spinor $\epsilon^{\mu_1,...\mu_{j-1/2}}_\alpha$ subject to the constraints of symmetry, transversality and tracelessness
$v_{\mu} \epsilon^{\mu,...\mu_{j-1/2}}_\alpha = 0$, $g_{\mu\nu}\ \epsilon^{\mu,\nu...\mu_{j-1/2}}_\alpha = 0$ and
$({/\hskip - 2 truemm v})_{\alpha\beta} \epsilon^{\mu_1,...,\mu_{j-1/2}}_\beta  = 0$, $(\gamma_{\mu_1})_{\alpha\beta} \epsilon^{\mu_1,...,\mu_{j-1/2}}_\beta = 0 \,,$
Then a light cloud scalar product
\begin{equation}
<v',j',\epsilon'|v,j,\epsilon>\,,
\label{ra_eq36}
\end{equation} 

\noindent that gives the IW function, is a covariant function of the vectors $v$ and $v'$ and of the tensors (or tensor-spinors) $\epsilon'^*$ and $\epsilon$, bilinear with respect to $\epsilon'^*$ and $\epsilon$, and the IW functions, functions of the scalar $v.v'$, are introduced accordingly.\par
The covariance property of the scalar products is explicitly expressed by the equality
\begin{equation}
<\Lambda v',j',\Lambda \epsilon'|\Lambda v,j,\Lambda \epsilon>\ =\ <v',j',\epsilon'|v,j,\epsilon>\,,
\label{ra_eq37}
\end{equation} 

\noindent valid for any Lorentz transformation $\Lambda$, with the transformation of a tensor-spinor given by
\begin{equation}
(\Lambda \epsilon)^{\mu_1,...,\mu_{j-1/2}}_\alpha = \Lambda^{\mu_1}_{\nu_1} ...  \Lambda^{\mu_{j-1/2}}_{\nu_{j-1/2}} D(\Lambda)_{\alpha\beta}\ \epsilon^{\nu_1,...,\nu_{j-1/2}}_\beta\,.
\label{ra_eq38}
\end{equation} 

Then, let us {\it define} the operator $U(\Lambda)$, in the space of the light cloud states, by
\begin{equation}
U(\Lambda)|v,j,\epsilon>\ = |\Lambda v,j,\Lambda\epsilon>\,,
\label{ra_eq39}
\end{equation}
 
\noindent where here $v$ is a fixed, arbitrarily chosen velocity. Eq. (\ref{ra_eq37}) implies that $U(\Lambda)$ is a {\it unitary operator}, as demonstrated in ref. \cite{LOR-3}.

A unitary representation of the Lorentz group emerges thus from the usual treatment of heavy hadrons in the heavy quark theory. For the present purpose, we need to go in the opposite way, namely, to show how, starting from a unitary representation of the Lorentz group, the usual treatment of heavy hadrons and the introduction of the IW functions emerges. What follows is not restricted to the $j = {1 \over 2} $ case, but concerns any IW function.\par
So, let us consider some unitary representation $\Lambda \to U(\Lambda)$ of the Lorentz group, or more precisely of the group $SL(2,C)$, in a Hilbert space $\mathcal{H}$, and we have to identify states in $\mathcal{H}$, depending on a velocity $v$. As said in \cite{LOR-3}, we have in $\mathcal{H}$ an additional structure, namely the energy operator of the light component {\it for a heavy quark at rest}, with $v_0=(1,0,0,0)$. Since this energy operator is invariant under rotations, we consider the subgroup $SU(2)$ of $SL(2,C)$. By restriction, the representation in $\mathcal{H}$ of $SL(2,C)$ gives a representation $R \to U(R)$ of $SU(2)$, and its decomposition into irreducible representations of $SU(2)$ is needed. We then have the eigenstates $|v_0,j,M>$ of the energy operator, classified by the angular momentum number $j$ of the irreducible representations of $SU(2)$, and {\it associated with the rest velocity} $v_0$, since their physical meaning is to describe the energy eigenstates of the light component for a heavy quark at rest.\par
We need now to express the states $|v,j,\epsilon>$ in terms of the states $|v_0,j,M>$. We begin with $v = v_0$. For fixed $j$ and $\alpha$, the states $|v_0,j,M>$ are, for $-j \leq M \leq j$, a standard basis of a representation $j$ of $SU(2)$ : 
\begin{equation} 
U(R)\ |v_0,j,M>\ = \sum_{M'}\ D^j_{M',M}(R)\ |v_0,j,M'>\,,
\label{ra_eq40}
\end{equation} 

\noindent where the rotation matrix elements $D^j_{M',M}$ are defined by
\begin{equation} 
D^j_{M',M} =\ <j,M'|U_j(R)|j,M>\,.
\label{ra_eq41}
\end{equation} 

On the other hand, the states $|v_0,j,\epsilon>$ constitute, when $\epsilon$ goes over all polarization tensors (or tensor-spinors), the whole space of a representation $j$ of $SU(2)$. As emphasized in \cite{LOR-3}, the representation of $SU(2)$ in the space of 3-tensors (or 3-tensor-spinors) is not irreducible, but contains the irreducible subspace of spin $j$, which is precisely the polarization 3-tensor (or 3-tensor-spinor) space selected by the other constraints.\par
We may then introduce a standard basis $\epsilon^{(M)}$, $-j \leq M \leq j$, for the $SU(2)$ representation of spin $j$ in the space of polarization 3-tensors (or 3-tensor-spinors). As demonstrated in \cite{LOR-3}, for any Lorentz transformation $\Lambda$ we must have
\begin{equation} 
|v,j,\epsilon>\ = \sum_{M} (\Lambda^{-1}\epsilon)_M\ U(\Lambda) |v_0,j,M>\,,
\label{ra_eq42}
\end{equation} 

\noindent for $\Lambda$ such that $\Lambda v_0 = v$, with $v_0 = (1,0,0,0)$.
 
Equation (\ref{ra_eq42}) is {\it our main result} here, defining, in the Hilbert space $\mathcal{H}$ of a unitary representation of $SL(2,C)$, the states $|v,j,\epsilon>$ whose scalar products define the IW functions, in terms of $|v_0,j,M>$ which occur as $SU(2)$ multiplets in the restriction to $SU(2)$ of the $SL(2,C)$ representation.\par

\section{Irreducible unitary representations of the Lorentz group and their decomposition under rotations} 

\subsection{Explicit form of the principal series of irreducible unitary representations of the Lorentz group}
Following Na\"{\i}mark \cite{NAIMARK}, we have exposed in \cite{LOR-3} an explicit form of the irreducible unitary representations of $SL(2,C)$. Their set $X$ is divided into three sets, the set $X_p$ of representations of the principal series, the set $X_s$ of representations of the supplementary series, and the one-element set $X_t$ made up of the trivial representation.
Actually, for the $j = {1 \over 2}$ case, only the principal series is relevant, and we now consider the principal series, leaving $j$ completely general. \par
A representation $\chi = (n,\rho)$ in the principal series is labelled by an integer $n \in Z$ and a real number $\rho \in R$. Actually, the representations $(n,\rho)$ and $(-n,-\rho)$ (as given below) turn out to be equivalent so that, in order to have each representation only once, $n$ and $\rho$ will be restricted as follows :
$$n = 0\,, \qquad\qquad\qquad \rho \geq 0\,,$$
\begin{equation} 
n > 0\,, \qquad\qquad\qquad \rho \in R \,.
\label{ra_eq43}
\end{equation} 

The Hilbert space $\mathcal{H}_{n,\rho}$ is made up of functions of a complex variable $z$ with the standard scalar product
\begin{equation}
<\phi'|\phi>\ = \int \overline{\phi'(z)}\ \phi(z)\ d^2z\,.
\label{ra_eq44}
\end{equation} 

\noindent with the measure $d^2z$ in the complex plane being simply $d^2z = d({\rm Re} z)d({\rm Im} z)$, and therefore $\mathcal{H}_{n,\rho} = L^2(C,d^2z)$.\par
The unitary operator $U_{n,\rho}(\Lambda)$ is given by :
\begin{equation}
\left(U_{n,\rho}(\Lambda)\phi \right)\!(z) =  \left({{\alpha-\gamma z} \over {|\alpha-\gamma z|}}\right)^n |\alpha-\gamma z|^{2i\rho-2}\ \phi\!\left({\delta z-\beta} \over {\alpha-\gamma z}\right)\,,
\label{ra_eq45}
\end{equation} 

\noindent where $\alpha$, $\beta$, $\gamma$, $\delta$ are complex matrix elements of $\Lambda \in SL(2,C)$ :
\begin{equation}
\Lambda = \left( \begin{array}{cc} \alpha & \beta \\ \gamma & \delta \end{array} \right)\,, \qquad\qquad\qquad \alpha \delta - \beta \gamma = 1\,.
\label{ra_eq46}
\end{equation} 

\subsection{Decomposition under the rotation group}
Next we need the decomposition of the restriction to the subgroup $SU(2)$ of each irreducible unitary representation of $SL(2,C)$.\par
Since $SU(2)$ is compact, the decomposition is by a direct sum so that, for each representation $\chi \in X$ we have an {\it orthonormal basis} $\phi^\chi_{j,M}$ of $\mathcal{H}_\chi$ adapted to $SU(2)$. Having in mind the usual notation for the spin of the light component of a heavy hadron, here we denote by $j$ the spin of an irreducible representation of $SU(2)$. It turns out \cite{LOR-3} that each representation $j$ of $SU(2)$ appears in $\chi$ with multiplicity 0 or 1, so that $\phi^\chi_{j,M}$ needs no more indices, and the values taken by $j$ are integer and half-integer numbers. For fixed $j$, the functions $\phi^\chi_{j,M}$, $-j \leq M \leq j$ are choosen as a standard basis of the representation $j$ of $SU(2)$. \par
It turns out \cite{LOR-3} that the functions $\phi^\chi_{j,M}(z)$ are expressed in terms of the rotation matrix elements $D^j_{M',M}$ defined by (\ref{ra_eq41}). A matrix $R \in SU(2)$ being of the form
\begin{equation}
R = \left( \begin{array}{cc} a & b \\ -\overline{b} & \overline{a} \end{array} \right), \qquad\qquad\qquad |a|^2+|b|^2 = 1\,.
\label{ra_eq47}
\end{equation}

\noindent we shall also consider $D^j_{M',M}$ as a function of $a$ and $b$, satisfying $|a|^2+|b|^2 = 1$.\par

We can now give explicit formulae for the orthonormal basis $\phi^\chi_{j,M}$ of $\mathcal{H}_\chi$.\par

The spins $j$ which appear in a representation $\chi = (n,\rho)$ are \cite{NAIMARK,LOR-3} :
\begin{equation}
{\rm all\ integers} \qquad \qquad \qquad \ \ j \geq {n \over 2} \qquad \ \ \  {\rm for} \qquad n \qquad {\rm even}\,,
 \qquad
\label{ra_eq48}
\end{equation}
\begin{equation}
{\rm all\ half-integers} \qquad \qquad j \geq {n \over 2} \qquad \ \  {\rm for} \qquad n \qquad {\rm odd}\,. \qquad
\label{ra_eq49}
\end{equation}

\noindent and such a spin appears with multiplicity 1.\par
 The basis functions $\phi^{n,\rho}_{j,M}(z)$ are given by the expression \cite{LOR-3}
\begin{equation}
\phi^{n,\rho}_{j,M}(z) =  {\sqrt{2j+1} \over \sqrt{\pi}}\ (1+|z|^2)^{i\rho-1} D^j_{n/2,M}\! \left( {1 \over \sqrt{1+|z|^2}}, - {z \over \sqrt{1+|z|^2}} \right)\,,
\label{ra_eq50}
\end{equation}	

\noindent or, using the explicit formula for $D^j_{n/2,M}$,
$$\phi^{n,\rho}_{j,M}(z) = {\sqrt{2j+1} \over \sqrt{\pi}}\ (-1)^{n/2-M}\sqrt{{(j-n/2)!(j+n/2)!} \over {(j-M)!(j+M)!}}\ (1+|z|^2)^{i\rho-j-1}$$ 
\begin{equation}
\sum_{k}\ (-1)^k  \left( \begin{array}{c} j+M \\ k \end{array} \right) \left( \begin{array}{c} j-M \\ j-n/2-k \end{array} \right) z^{n/2-M+k}\ \overline{z}^k\,, 
\label{ra_eq51}
\end{equation}	

\noindent where the range for $k$ is limited to $0 \leq k \leq{j- {n \over 2}}$ due to the binomial factors.\par

\section{Irreducible Isgur-Wise functions for $j = {1 \over 2}$}

We are interested now in the ground state meson case $j = {1 \over 2}$ \cite{LOR-4}, for which from (\ref{ra_eq43})(\ref{ra_eq49}) one has a fixed value for $n$
\begin{equation}
j = {1 \over 2} \qquad \Rightarrow \qquad n = 1\,, \qquad \rho \in R\,.
\label{ra_eq52}
\end{equation}	

Deleting from now on  the fixed indices $j = {1 \over 2}$ and $n = 1$, and particularizing the explicit formula (\ref{ra_eq51}) to this case, we have :   
\begin{equation}
\phi^{\rho}_{+ {1 \over 2}}(z) = \sqrt{2 \over \pi} \left(1 + |z|^2 \right)^{i\rho - {3 \over 2}}\,, \qquad \qquad
\phi^{\rho}_{- {1 \over 2}}(z) = -\sqrt{2 \over \pi}\ z\left(1 + |z|^2 \right)^{i\rho - {3 \over 2}}\,.
\label{ra_eq53}
\end{equation}	

Let us now particularize the $SL(2,C)$ matrix (\ref{ra_eq46}) to a boost in the $z$ direction :
\begin{equation}
\Lambda_\tau = \left( \begin{array}{cc} e^{\tau \over 2} & 0 \\ 0 & e^{- {\tau \over 2}} \end{array} \right)\,, \qquad \qquad \qquad w = \cosh(\tau)\,.
\label{ra_eq54}
\end{equation}	

\noindent and following the $j = 0$ case studied at length in ref. \cite{LOR-3} let us consider the objects
\begin{equation}
\xi_\rho^{\pm {1 \over 2}, \pm {1 \over 2}}(w) =\ <\phi^\rho_{\pm {1 \over 2}}|U^\rho(\Lambda_\tau)\phi^\rho_{\pm {1 \over 2}}>\,.
\label{ra_eq55}
\end{equation}	

\noindent From the transformation law (\ref{ra_eq45}) and the explicit form (\ref{ra_eq53}), one gets :
$$\left(U^\rho(\Lambda_\tau)\phi^\rho_{+ {1 \over 2}}\right)(z) = \sqrt{2 \over \pi}\ e^{(i\rho-1)\tau}\left(1 + e^{-2\tau}|z|^2 \right)^{i\rho - {3 \over 2}}\,,$$
\begin{equation}
\left(U^\rho(\Lambda_\tau)\phi^\rho_{- {1 \over 2}}\right)(z) = - \sqrt{2 \over \pi}\ e^{(i\rho-1)\tau} e^{-\tau} z \left(1 + e^{-2\tau}|z|^2 \right)^{i\rho - {3 \over 2}}\,.
\label{ra_eq56}
\end{equation}	

\noindent and from these expressions one obtains :
$$\xi_\rho^{+ {1 \over 2}, + {1 \over 2}}(w) = {2 \over \pi} \int \left(1 + |z|^2 \right)^{-i\rho - {3 \over 2}} e^{(i\rho-1)\tau} \left(1 + e^{-2\tau}|z|^2 \right)^{i\rho - {3 \over 2}} d^2z\,, \qquad \qquad$$
\begin{equation}
\xi_\rho^{- {1 \over 2}, - {1 \over 2}}(w) = {2 \over \pi} \int e^{-\tau} |z|^2 \left(1 + |z|^2 \right)^{-i\rho - {3 \over 2}} e^{(i\rho-1)\tau} \left(1 + e^{-2\tau}|z|^2 \right)^{i\rho - {3 \over 2}} d^2z\,. \ \ \
\label{ra_eq57}
\end{equation}	

We must now extract the Lorentz invariant Isgur-Wise function $\xi(w)$. To do that, we must decompose into invariants the matrix elements (\ref{ra_eq57}) using the spin ${1 \over 2}$ spinors of the light cloud $u_{\pm {1 \over 2}}$. Since we have not introduced parity in our formalism we will have the following decomposition :
\begin{equation}
\xi_\rho^{\pm {1 \over 2}, \pm {1 \over 2}}(w) = \left(\overline{u}_{\pm {1 \over 2}}(v')u_{\pm {1 \over 2}}(v) \right) \xi^\rho(w) + \left(\overline{u}_{\pm {1 \over 2}}(v')\gamma_5u_{\pm {1 \over 2}}(v) \right) \tau^\rho(w)\,,
\label{ra_eq58}
\end{equation}	

\noindent where $\xi^\rho(w)$ is an irreducible ${1 \over 2}^- \to {1 \over 2}^-$ elastic IW function, labelled by the index $\rho$, and $\tau^\rho(w)$ is a function corresponding to the flip of parity ${1 \over 2}^- \to {1 \over 2}^+$.\par
One gets for the spinor bilinear
\begin{equation}
\overline{u}_{+ {1 \over 2}}(v')u_{+ {1 \over 2}}(v) = \overline{u}_{- {1 \over 2}}(v')u_{- {1 \over 2}}(v) = \sqrt{w+1 \over 2}\,,
\label{ra_eq59}
\end{equation}	

\noindent and since $\overline{u}_{+ {1 \over 2}}(v')\gamma_5 u_{+ {1 \over 2}}(v) = - \overline{u}_{- {1 \over 2}}(v') \gamma_5 u_{- {1 \over 2}}(v)$,
\begin{equation}
\xi_\rho(w) = \sqrt{2 \over w+1}\ {1 \over 2}\ \left[\xi^\rho_{+ {1 \over 2}, + {1 \over 2}}(w) + \xi^\rho_{- {1 \over 2}, - {1 \over 2}}(w) \right]\,,
\label{ra_eq60}
\end{equation}	

\noindent one obtains finally :
\begin{equation}
\xi_\rho(w) = {1 \over 1+\cosh(\tau)}{1 \over \sinh(\tau)}\ {4 \over 4\rho^2+1}\left[\sinh\left({\tau \over 2} \right)\cos(\rho\tau)+2\rho\cosh\left({\tau \over 2}\right)\sin(\rho\tau)\right]\,.
\label{ra_eq61}
\end{equation}	

This is the expression for the elastic ${1 \over 2}^- \to {1 \over 2}^-$ {\it irreducible Isgur-Wise functions} we were looking for, parametrized by the real parameter $\rho$ that labels the irreducible representations. The irreducible IW functions satisfy 
\begin{equation}
\xi_\rho(1) = 1\,.
\label{ra_eq62}
\end{equation}	

Like in the case $j = 0$, analized in great detail in \cite{LOR-3}, the elastic ${1 \over 2}^- \to {1 \over 2}^-$ IW function $\xi(w)$ will be given by the integral over a positive measure $d\nu(\rho)$ :
\begin{equation}
\xi(w) = \int_{]-\infty,\infty[} \xi_\rho(w)\ d\nu(\rho)\,,
\label{ra_eq63}
\end{equation}	

\noindent where the measure is normalized acording to
\begin{equation}
\int_{]-\infty,\infty[} d\nu(\rho) = 1\,.
\label{ra_eq64}
\end{equation}	

\noindent Notice that the range for the parameter $\rho$ that labels the irreducible representations follows from the fact that in the $j = {1 \over 2}$ case one has $n = 1$ and $\rho \in R$, eq. (\ref{ra_eq52}). Notice also that the IW irreducible function (\ref{ra_eq61}) is even in $\rho$, $\xi_\rho(w) = \xi_{-\rho}(w)$.\par
The irreducible IW functions (\ref{ra_eq61}), parametrized by some fixed value of $\rho = \rho_0$, are legitimate IW functions since the corresponding measure is given by a delta function,
\begin{equation}
d\nu(\rho) = \delta(\rho-\rho_0)\ d\rho\,.
\label{ra_eq65}
\end{equation}	

In the case of the irreducible representation $\rho_0 = 0$ one finds
\begin{equation}
\xi_0(w) = {4 \sinh\left({\tau \over 2}\right) \over (1+\cosh(\tau))\sinh(\tau)} = \left({2 \over 1+w}\right)^{3 \over 2}\,,
\label{ra_eq66}
\end{equation}	

\noindent that saturates the lower bound for the slope $-\xi'(1) \geq {3 \over 4}$. This is the so-called BPS limit of the IW function, considered in Section 4 within the Sum Rule approach, where we have seen that in the limit $-\xi'(1) \to {3 \over 4}$ one obtains $\xi(w) \to \xi_0(w)$.

\section{Integral formula for the IW function $\xi(w)$ and polynomial expression for its derivatives} 

From the norm and the normalization of the irreducible IW functions one gets the correct value of the IW function at zero recoil $\xi(1) = 1$. The integral formula writes, explicitly,
$$\xi(w) = {1 \over 1+\cosh(\tau)}{1 \over \sinh(\tau)}$$
\begin{equation}
\times  \int_{]-\infty,\infty[}{4 \over 4\rho^2+1} \left[\sinh\left({\tau \over 2} \right)\cos(\rho\tau)+2\rho\cosh\left({\tau \over 2}\right)\sin(\rho\tau)\right]\ d\nu(\rho)\,.
\label{ra_eq67}
\end{equation}	

\noindent from which one can find the following polynomial expression for its derivatives :
\begin{equation}
\xi^{(n)}(1) = (-1)^n {1 \over 2^{2n}(2n+1)!!} \prod_{i=1}^{n} \left< \left[ (2i+1)^2+4\rho^2 \right] \right> \qquad \qquad (n \ge 1)\,.
\label{ra_eq68}
\end{equation}

\noindent where the mean value is defined as $\left< f(\rho) \right> = \int_{]-\infty,\infty[} f(\rho)\ d\nu(\rho)$. This formula can be demonstrated along the same lines as the corresponding one in the baryon case done in Appendix D of ref. \cite{LOR-3}.

\section{Lower bounds on the derivatives of the IW function} 

From (\ref{ra_eq68}) one gets immediately the lowest bounds on the derivatives (\ref{ra_eq24}) obtained using the SR approach.

To get improved bounds on the derivatives we must, like in \cite{LOR-3}, express the derivatives in terms of moments of the {\it positive} variable $\rho^2$, that can be read from (\ref{ra_eq68}). Calling the moments : 
\begin{equation}
\mu_n =\ <\rho^{2n}>\ \ge\ 0 \qquad \qquad (n \ge 0)\,,
\label{ra_eq69}
\end{equation}

\noindent one gets the successive derivatives in terms of moments :
\begin{equation}
\xi(1) = \mu_0 = 1, \qquad \xi'(1) = - \left({3 \over 4} + {1 \over 3} \mu_1 \right), \qquad \xi''(1) = {15 \over 16} + {17 \over 30} \mu_1 + {1 \over 15} \mu_2,...
\label{ra_eq70}
\end{equation}

The relations (\ref{ra_eq70}) can be solved step by step, and the moment $\mu_n$ is expressed as a combination of the derivatives $\xi(1)$, $\xi'(1)$,... $\xi^{(n)}(1)$ :
\begin{equation}
\mu_0 = \xi(1) = 1, \ \ \ \mu_1 = - {3 \over 4} \left[ 3+4\xi'(1) \right],  \ \ \ \mu_2 = {3 \over 16} \left[ 27+136\xi'(1)+80\xi''(1) \right],...
\label{ra_eq71}
\end{equation}

Since $\rho^2$ is a positive variable, one can obtain improved bounds on the derivatives from the following set of constraints. For any $n \geq 0$, one has \cite{LOR-3}
\begin{equation}
\det\left[(\mu_{i+j})_{0\leq i,j \leq n}\right] \geq 0, \qquad \qquad \det\left[(\mu_{i+j+1})_{0\leq i,j \leq n}\right] \geq 0\,.
\label{ra_eq72}
\end{equation}

Since each moment $\mu_k$ is a combination of the derivatives $\xi(1)$, $\xi'(1)$,... $\xi^{(k)}(1)$, the constraints on the moments translate into constraints on the derivatives. Using (\ref{ra_eq72}) one gets positivity conditions of the form
\begin{equation} 
\mu_1 \geq 0, \qquad \qquad \det \left( \begin{array}{cc}1&\mu_1\\\mu_1&\mu_2\\\end{array} \right) = \mu_2 - \mu_1^2 \geq 0,...
\label{ra_eq73}
\end{equation} 

\noindent that imply 
\begin{equation} 
\mu_1 \geq 0, \qquad \qquad \mu_2 \geq \mu_1^2,...
\label{ra_eq74}
\end{equation} 

These constraints imply, in terms of the derivatives :
\begin{equation} 
-\xi'(1) \geq {3 \over 4}, \qquad \qquad \xi''(1) \geq {1 \over 5}\left[-4\xi'(1)+3\xi'(1)^2\right],... 
\label{ra_eq75}
\end{equation} 

\noindent We see that we recover the bounds obtained using the SR method. The method generalizes in a straightforward way to higer derivatives.\par

\section{Inversion of the integral representation of the IW function}

Let us now show that the integral formula for the IW function (\ref{ra_eq67}) can be inverted, giving the positive measure $d\nu(\rho)$ in terms of the IW function $\xi(w)$. This will allow to formulate criteria to test the validity of a given phenomenological ansatz for $\xi(w)$.\par
Let us define
\begin{equation} 
\widehat{\xi}(\tau) = (\cosh(\tau)+1)\sinh(\tau)\xi(\cosh(\tau))\,.
\label{ra_eq76}
\end{equation} 

\noindent and similarly for $\widehat{\xi}_\rho(\tau)$. 

The integral formula then writes
\begin{equation} 
\widehat{\xi}(\tau) = \int \widehat{\xi}_\rho(\tau)d\nu(\rho) = \int  (\cosh(\tau)+1)\sinh(\tau) \xi_\rho(\cosh(\tau))d\nu(\rho)\,.
\label{ra_eq77}
\end{equation} 

One finds, for its derivative, the simple expression :
\begin{equation} 
{d \over d\tau}\ \widehat{\xi}_\rho(\tau) = 2 \cos(\rho\tau)\cosh\left({\tau \over 2}\right)\,.
\label{ra_eq78}
\end{equation} 

Defining the function
\begin{equation} 
\eta(\tau) = {1 \over 2\cosh\left({\tau \over 2}\right)} {d \over d\tau}\ \widehat{\xi}(\tau)\,,
\label{ra_eq79}
\end{equation}

\noindent one sees that the integral formula reads simply
\begin{equation} 
\eta(\tau) = \int_{]-\infty,\infty[} \cos(\rho \tau) d\nu(\rho)\,,
\label{ra_eq80}
\end{equation}

Computing the Fourier transform
\begin{equation} 
{\tilde \eta}(\rho) = {1 \over 2\pi}\int_{-\infty}^{+\infty} e^{i\tau\rho}\ d\tau\ \eta(\tau) = \int_{]-\infty,\infty[} {1 \over 2}\ [\delta(\rho'+\rho)+\delta(\rho'-\rho)]\ d\nu(\rho')\,,
\label{ra_eq81}
\end{equation}

\noindent and defining the function
\begin{equation} 
\mu(\rho) = {d\nu(\rho) \over d\rho}\,,
\label{ra_eq82}
\end{equation}

\noindent one finds
\begin{equation} 
{\tilde \eta}(\rho) = {1 \over 2}\ [\mu(\rho)+\mu(-\rho)]\,. 
\label{ra_eq83}
\end{equation}

We now assume that the general measure $d\nu(\rho)$ is even, i.e. it has the same parity as the measure $d\rho$, without loss of generality because $\xi_\rho(w)$ is even in $\rho$.\par 
Then, the function (\ref{ra_eq82}) is even $\mu(\rho) = \mu(-\rho)$ and one finally finds for the measure $d\nu(\rho) = {\tilde \eta}(\rho) d\rho$ : 
\begin{equation} 
d\nu(\rho) = {1 \over 2\pi}\int_{-\infty}^{+\infty} e^{i\tau\rho}\ d\tau\ {1 \over 2\cosh\left({\tau \over 2}\right)} {d \over d\tau} \left[(\cosh(\tau)+1)\sinh(\tau)\xi(\cosh(\tau))\right]\ d\rho\,.
\label{ra_eq84}
\end{equation} 

This completes the inversion of the integral representation. Equation (\ref{ra_eq84}) is the master formula expressing the measure in terms of the Isgur-Wise function.\par
One can apply this formula to check if a given phenomenological formula for the IW function $\xi(w)$ satisfies the constraint that the corresponding measure $d\nu(\rho)$ must be positive. This provides a powerful consistency test for any proposed ansatz.

\section{An upper bound on the Isgur-Wise function}

From the integral formula also an upper bound on the whole IW function $\xi(w)$ can be obtained.
Defining the function
\begin{equation}
\eta_\rho(\tau) = {1 \over 2\cosh\left({\tau \over 2}\right)} {d \over d\tau}\ \widehat{\xi}_\rho(\tau)\,,
\label{ra_eq85}
\end{equation} 

\noindent we have obtained above : 
\begin{equation}
\eta_\rho(\tau) =\cos(\rho \tau)\,,
\label{ra_eq86}
\end{equation}

\noindent and from it it follows
\begin{equation}
-1 \leq \eta_\rho(\tau) \leq 1\,,
\label{ra_eq87}
\end{equation}

\noindent and hence
\begin{equation}
-2 \cosh\left({\tau \over 2}\right) \leq {d \over d\tau}\ \widehat{\xi}_\rho(\tau) \leq 2 \cosh\left({\tau \over 2}\right)\,.
\label{ra_eq88}
\end{equation}

\noindent Integrating this inequality from $0$, one gets :
\begin{equation}
- 4 \sinh\left({\tau \over 2}\right) \leq \widehat{\xi}_\rho(\tau) \leq 4 \sinh\left({\tau \over 2}\right)\,,
\label{ra_eq89}
\end{equation}

\noindent and since
\begin{equation}
\widehat{\xi}_0(\tau) = 4 \sinh\left({\tau \over 2}\right)\,,
\label{ra_eq90}
\end{equation}

\noindent one finds the inequalities 
\begin{equation}
- \widehat{\xi}_0(\tau) \leq \widehat{\xi}_\rho(\tau) \leq \widehat{\xi}_0(\tau)\,,
\label{ra_eq91}
\end{equation}

\noindent that simplify to :
\begin{equation}
- \xi_0(\tau) \leq \xi(\tau) \leq \xi_0(\tau)\,.
\label{ra_eq92}
\end{equation}

Since $\xi_0(\tau)$ is given by the expression (\ref{ra_eq66}), one finally obtains
\begin{equation}
\left\vert \xi(w) \right\vert \ \leq  \left({2 \over 1+w}\right)^{3 \over 2}\,.
\label{ra_eq93}
\end{equation}

\noindent This inequality is a strong result because it holds for any value of $w$.

\section{Consistency tests for any ansatz of the Isgur-Wise function : phenomenological applications}

To  illustrate the methods exposed above, we now examine some phenomenological formulas proposed in the literature. In ref.  \cite{LOR-4} we have studied a number of other interesting cases.\par 
We will compare these ansatze with the theoretical criteria that we have formulated : the lower bounds on the derivatives at zero recoil, the upper bound obtained for the whole IW function, and the inversion of the integral formula for the IW function in order to check the positivity of the measure.\par
We must underline that the satisfaction of the bounds on the derivatives and of the upper bound on the whole IW function are {\it necessary conditions}, while the criterium of the positivity of the measure is a {\it necessary and sufficient condition} to establish if a given ansatz of the IW function satisfies the Lorentz group criteria.

\subsection{The exponential ansatz}

This form corresponds to the non-relativistic limit for the light quark with the harmonic oscillator potential \cite{JLOR-2} :
\begin{equation}
\xi(w) = \exp\left[-c(w-1)\right]\,.
\label{ra_eq94}
\end{equation} 

The bound for the slope is satisfied for $c \geq {3 \over 4}$, the bound for the second derivative is satisfied for $c \geq 2$, while the bound for the third derivative is violated for any value of c.
Therefore, this phenomenological ansatz on the IW function is invalid.\par
The exponential ansatz satisfies nevertheless the upper bound (\ref{ra_eq93}).\par

Let us now examine the criterium based on the positivity of the measure. One needs to compute
\begin{equation}
\eta(\tau) = {1 \over c} \left(- {d^2 \over d\tau^2} + {1 \over 4} \right) \cosh\left({\tau \over 2}\right)\exp\left[-c(\cosh(\tau)-1)\right]\,.
\label{ra_eq95}
\end{equation} 

\noindent The function $\eta(\tau)$ is bounded for any value of $c$. The Fourier transform of this function gives 
\begin{equation}
d\nu(\rho) = {e^c \over 2\pi} {1 \over c} \left( \rho^2 + {1 \over 4} \right) \left[ K_{i\rho + {1 \over 2}}(\rho) + K_{-i\rho + {1 \over 2}}(\rho) \right] d\rho\,.
\label{ra_eq96}
\end{equation} 

\noindent Since this function is not positive for any value of $c$, the exponential ansatz for the IW function violates the consistency criteria exposed above.

\subsection{The "dipole"}

The following shape has been proposed in the literature (see for example \cite{NRSX,MLOPR-2})
\begin{equation}
\xi(w) = \left({2 \over 1+w}\right)^{2c}\,.
\label{ra_eq97}
\end{equation} 

\noindent The bounds for the slope and for the higher derivatives are satisfied for $c \geq {3 \over 4}$.\par

Let us now compute the measure (\ref{ra_eq84}).
One needs first to compute
\begin{equation}
\eta(\tau) = -4(c-1) \left[\cosh\left({\tau \over 2}\right) \right]^{-4c+3} + (4c-3) \left[\cosh\left({\tau \over 2}\right)\right]^{-4c+1}\,.
\label{ra_eq98}
\end{equation} 

\noindent Since $\eta(\tau)$ has to be bounded, the parameter $c$ must satisfy $c \geq {3 \over 4}$.

Moreover, we realize that in the particular case
\begin{equation}
c = {3 \over 4} \qquad \ \ \to \qquad \ \ \eta(\tau) = 1 \qquad \ \ \to \qquad \ \ d\nu(\rho) = \delta(\rho)\ d\rho\,.
\label{ra_eq99}
\end{equation} 

\noindent Therefore, one gets in this case a delta-function for the measure. This is a positive measure that corresponds to the explicit formula (\ref{ra_eq66}) for the IW function in the BPS limit.\par

For $c > {3 \over 4}$ one obtains a function $\eta(\tau)$ that is bounded and integrable. Computing its Fourier transform one gets the measure  
\begin{equation}
d\nu(\rho) = {2^{4c-1} \over 2\pi}\ (4c-3) \left( \rho^2 + {1 \over 4} \right){\Gamma\left(i\rho+2c-{3 \over 2}\right)\Gamma\left(-i\rho+2c-{3 \over 2}\right) \over \Gamma\left(4c-1\right)} \ d\rho\,,
\label{ra_eq100}
\end{equation} 

\noindent that is positive.\par
In conclusion, the measure $d\nu(\rho)$ for the "dipole" ansatz is positive for $c \geq {3 \over 4}$ and therefore satisfies all the consistency criteria.

\section{Conclusion}

We have reviewed a number of important works by Nikolai Uraltsev on Sum Rules in the heavy quark limit of QCD.\par
We have generalized Bjorken and Uraltsev SR to higher derivatives and we have formulated lower bounds on the successive derivatives of the Isgur-Wise function $\xi(w)$. We have also obtained an explicit form for the IW function in the "BPS limit" considered by N. Uraltsev.\par
On the other hand, the Lorentz group acting on the light cloud provides a transparent physical interpretation of the results obtained from the SR. Both methods are completely equivalent.\par
Within the Lorentz group method we have obtained an integral formula for the IW function in terms of an explicit kernel and {\it a positive measure}.\par
 From this representation we have reproduced the bounds on the derivatives of the IW function from positivity conditions on moments of a positive variable.\par
On the other hand, we have inverted the integral formula expressing the positive measure in terms of any given ansatz of the IW function.\par
As a consence, the "BPS limit" for the IW function obtained from the SR method turns out to have a clear group theoretical interpretation : the positive measure is just a $\delta$-function, and the cloud ground state belongs to a particular irreducible representation of $SL(2,C)$.
 
Some phenomenological proposals for the shape of the Isgur-Wise function have been compared with the theoretical constraints obtained in this paper.\par
These different shapes provide illustrations of the method in a rather complete way. The different criteria based on the Lorentz group, i.e. lower limits on the derivatives at zero recoil, positivity of the measure in the inversion formula for the IW function and the upper bound for the whole IW function, have been illustrated.\par
A main conclusion is that, using a method based on the Lorentz group, completely equivalent to the one of generalized Bjorken-Uraltsev sum rules, one obtains strong constraints on the Isgur-Wise function for the ground state mesons.

\section*{Acknowledgements}

We are indebted to our colleague Alain Le Yaouanc for discussions on  our common work and on the present text.

\end{document}